\documentclass[%
reprint,
superscriptaddress,
 amsmath,amssymb,
 aps,
pra,
]{revtex4-2}

\usepackage{graphicx}% Include figure files
\usepackage{dcolumn}% Align table columns on decimal point
\usepackage{bm}% bold math
\usepackage[hidelinks]{hyperref}
\usepackage{color}
\usepackage[export]{adjustbox}
\begin{document}

\preprint{APS/123-QED}
%%%%%
% Alternative words 
% IBM Quantum (IBM-Q) + processor, computer, device, system 
%%%%%
%\title{Error-disturbance uncertainty relations in IBM-Q processor}
\title{Error-disturbance uncertainty relations in a superconducting quantum processor}
%\title{Error-disturbance uncertainty relations in a superconducting quantum computer}
%\title{Error-disturbance uncertainty relations using a superconducting qubit on IBM-Q processor}
\author{Tingrui Dong}
\email{dong.tingrui.r7@dc.tohoku.ac.jp}
\affiliation{Research Institute of Electrical Communication, Tohoku University, Sendai 980-8577, Japan}
\affiliation{Department of Electronic Engineering, Graduate School of Engineering, Tohoku University, Sendai 980-8579, Japan}
\author{Soyoung Baek}
\email{soyoungb@tohoku.ac.jp}
\affiliation{Research Institute of Electrical Communication, Tohoku University, Sendai 980-8577, Japan}
\author{Fumihiro Kaneda}
\affiliation{Department of Physics, Graduate School of Science, Tohoku University, Sendai 980-8579 Japan}
\author{Keiichi Edamatsu}
\affiliation{Research Institute of Electrical Communication, Tohoku University, Sendai 980-8577, Japan}

\date{\today}

\begin{abstract}
We experimentally test the error-disturbance uncertainty relation (EDR) in generalized, variable strength measurements of superconducting qubits on a NISQ processor. 
Making use of sequential weak measurements that keeps the initial signal state practically unchanged prior to the main measurement, we demonstrate that the Heisenberg EDR is violated, yet the Ozawa and Branciard EDRs are valid throughout the range of measurement strengths from no measurement to projection measurement. 
Our results verify that universal EDRs are valid even in a noisy quantum processor and will stimulate research on measurement-based quantum information and communication protocols using a NISQ processor.
\end{abstract}

\maketitle
\section{Introduction\label{sec:introduction}}

%%%%%%%%%%%%%%
% Importance of EDR
%%%%%%%%%%%%%%
The error-disturbance uncertainty relation (EDR) is a cornerstone of quantum mechanics since it describes the fundamental limitations on the achievable accuracy of quantum measurement. 
Furthermore, with the fast development of quantum technologies such as quantum metrology \cite{10.1038/nphoton.2013.150}, quantum communications \cite{10.1007/3-540-53862-3_161}, and quantum computations \cite{nielsen_chuang_2010}, it has become more crucial for us to understand the ultimate quantum limitations in the attainable accuracy of quantum measurement. 

In 1927, Heisenberg  \cite{heisenberg1927} formulated that any measurement of the particle’s position $x$ with the error $\epsilon(x)$ causes the disturbance $\eta(p)$ on momentum $p$ satisfying $\epsilon(x)\eta(p)\geq\hbar/2$. The generalized form of Heisenberg’s EDR for an arbitrary pair of observables $A$ and $B$ is given by  
%%%%%%%%%%
\begin{equation}
\label{H_edr}
\epsilon(A)\eta(B)\geq C,
\end{equation}
%%%%%%%%%%
where the lower bound $C$ is determined by the commutation relation between $A$ and $B$, i.e., $C = | \langle [ \, A, B ] \,\rangle | /2$, $ [ \, A, B] \,=AB-BA$, and $ \langle ... \rangle$ stands for the mean value in a given state.

In 2003, Ozawa \cite{PhysRevA.67.042105} proposed an alternative EDR that is theoretically proven to be universally valid 
%%%%%%%%%%
\begin{equation}
\label{O_edr}
\epsilon(A)\sigma(B) + \sigma(A)\epsilon(B) +\epsilon(A)\eta(B) \geq C,
\end{equation}
%%%%%%%%%%
where $\sigma(A) = \sqrt{\langle A^2 \rangle - \langle A \rangle ^2}$ is the standard deviation. 
Ozawa’s relation has two additional correlation terms, the presence of which allows the error-disturbance product $\epsilon(x)\eta(p)$ to be below the lower bound of Eq. (\ref{H_edr}), implying the violation of Heisenberg's EDR.
Note that there are alternative approaches for the measurement uncertainty relations based on different definitions of error and disturbance \cite{Edamatsu_2016, PhysRevA.68.032103, PhysRevA.84.042121, PhysRevLett.111.160405, COLES2015105, PhysRevLett.115.030401}. 
For example, Busch \textit{et al.} \cite{PhysRevA.89.012129} defined measurement error based on the RMS distance of the distributions between the original and the measurement observables and derived EDR for the qubit case. 

Most recently, Branciard \cite{branciard2013error} has improved the Ozawa’s EDR and obtained even tighter EDR such as 
%%%%%%%%%%
\begin{equation}
\label{B_edr}
\begin{aligned}
& \left[ \epsilon(A)^2\sigma(B)^2+\sigma(A)^2\eta(B)^2   \right. \\
 &\left. + 2\epsilon(A)\eta(B) \sqrt{\sigma(A)^2 \sigma(B)^2- C^2} \right]^{1/2}  \geq C.
\end{aligned} 
\end{equation}
%%%%%%%%%%
Due to its fundamental and practical importance, EDRs have been investigated in various quantum systems such as polarized neutrons \cite{Erhart2012, PhysRevA.88.022110}, photons \cite{10.1103/physrevlett.109.100404, Baek2013, PhysRevLett.110.220402, PhysRevLett.112.020401, PhysRevLett.112.020402}, and trapped ions \cite{doi:10.1126/sciadv.1600578}.
% amplitude and phase quadratures of two EPR modes. 
Experimental tests of EDRs commonly proved that Heisenberg’s EDR can be violated in some systems, yet Ozawa’s and Branciard’s relations are always validated. 
 
%%%%%%%%%%%%%%
% Importance of NISQ processor
%%%%%%%%%%%%%%
Meanwhile, recent noisy intermediate-scale quantum (NISQ) technology is regarded as a step toward more powerful quantum technologies to be developed in the future \cite{Preskill2018quantumcomputingin}. 
Although NISQ devices are not capable of fault-tolerant operations, NISQ technology gives us new tools for exploring the physics of many entangled particles by allowing consecutive entangling operations between neighboring qubits. 
Proof-of-principle demonstrations of new quantum protocols using a NISQ processor also proves its prospect as a testbed of new quantum tasks \cite{PRXQuantum.3.040333, 10.1364/cleo_qels.2022.ff2i.1, tan2023extending}.

In this paper, we focus on availability of a NISQ processor as a quantum system to verify quantum measurement protocols. 
To prove it, we experimentally test the EDRs in generalized, variable strength measurements of superconducting qubits on the IBM Quantum (IBM-Q) processor \cite{ibm2021}. 
Making use of sequential weak measurements that keeps the initial signal state practically unchanged prior to the main measurement, we demonstrate that the Heisenberg EDR is violated, yet the Ozawa and Branciard EDRs are valid throughout the range of the measurement strength.
To the best of our knowledge, this is the first investigations of EDRs in a NISQ processor and our results verify that universal EDRs are valid even in noisy quantum processor. We anticipated that our study will stimulate research on various quantum measurement protocols using a quantum processor.

%%%%%%%%%%%%%%%%%%%%%%%%%%
\section{Scheme\label{sec:scheme}}
%%%%%%%%%%%%%%%%%%%%%%%%%%
%%%%%%%%%%%%%%%%%%%%%%%%%%
\begin{figure}[t] %fig1
\centering
\includegraphics[width=0.5 \textwidth, inner]{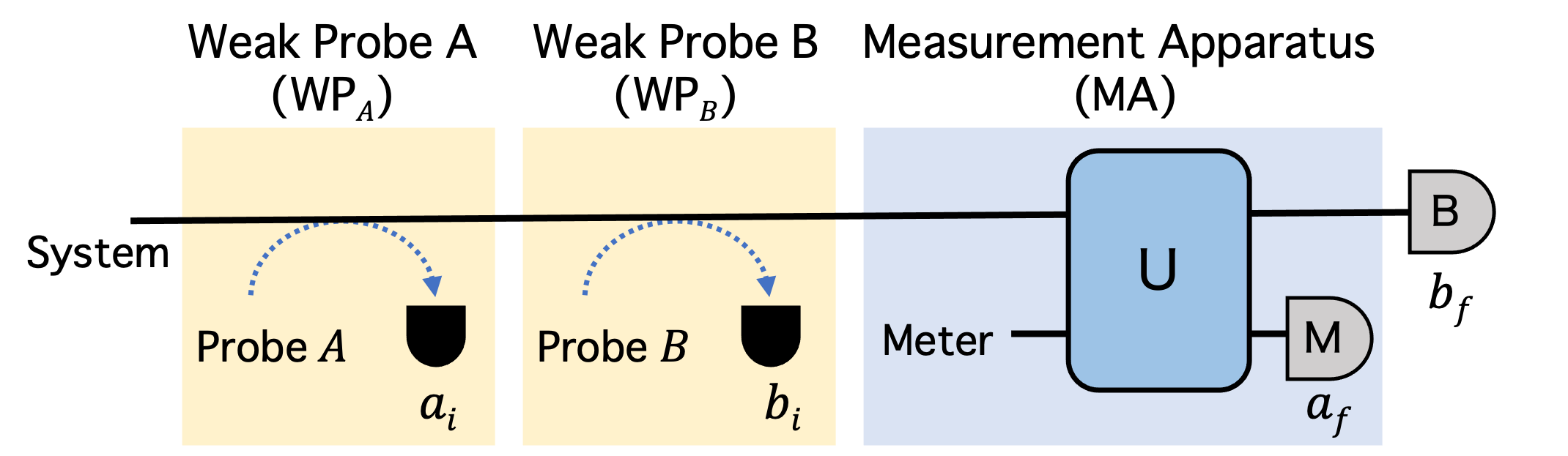} 
\caption{Schematic diagram to test EDRs using the weak-probe method for single-qubit observables $A$ and $B$. 
Weak probes indirectly measure $A$ and $B$ with a weak measurement strength, denoted as weak probe $A$ (WP$_A$) and  weak probe $B$ (WP$_B$), prior to the main measurement operated by the measurement apparatus (MA).We use sufficiently weak measurement strength for the WP that causes very little disturbance on the system state.} 
\label{scheme} 
\end{figure}
%%%%%%%%%%%%%%%%%%%%%%%%%%
%%%%%%%%%%%%%%
% Weak Probe Method (general)
%%%%%%%%%%%%%% 
The error and disturbance used to derive EDRs are defined from indirect measurement model, depicted as a ``measurement apparatus'' (MA) in Fig. \ref{scheme}. 
In that model, measurement of the system state is implemented by the interaction that (partially) correlates system and meter states, denoted by $U$, followed by the measurement of the meter states. 
Using stronger interaction, one can extract more information from the system state at the cost of a larger disturbance of the system state.
From the measurement model, the error and disturbance are defined as \cite{PhysRevA.67.042105}
%%%%%%%%%%
\begin{equation}\label{error_disturbance}
    \begin{aligned}   
    \epsilon (A) &= \langle [ U^\dagger (I \otimes M) U  - A\otimes I)] ^2\rangle^{1/2}, \\
    \eta (B) &= \langle [ U^\dagger (B \otimes I) U  - B\otimes I)] ^2\rangle^{1/2} ,
    \end{aligned}
\end{equation}
%%%%%%%%%%
where the average is taken over the system-meter composite state on input. 
$U$ is a unitary operator that provides interaction between the system and meter states, and $M$ is the meter observable.
The definition of $\epsilon(A)$ is uniquely derived from the classical notion of root-mean-square error if $U^\dagger (I \otimes M) U$ and $ A\otimes I$ commute, and otherwise, it is considered as a natural quantization of the notion of classical root-mean-square error \cite{PhysRevLett.112.020402}. 
The definition of $\eta(B)$ is derived analogously.

For experimental tests of EDRs, we employ the ``weak-probe method''\cite{Lund_2010, OZAWA200511} which has been used in the successful investigation of EDRs on single-photon polarizations by our co-authors \cite{PhysRevLett.112.020402}. 
Fig. \ref{scheme} shows the schematic diagram to test EDRs using the weak-probe method for single-qubit observables $A$ and $B$.
Weak probes indirectly measure $A$ and $B$ with a weak measurement strength, denoted as weak probe $A$ (WP$_A$) and  weak probe B (WP$_B$), prior to the main measurement operated by the measurement apparatus (MA).
When the measurement strength is sufficiently small, the system state is sent to the MA without being disturbed by the WP. 
As Lund and Wiseman \cite{Lund_2010}, and Ozawa \cite{OZAWA200511} pointed out, the error (disturbance) defined by Eq. (\ref{error_disturbance}) is given by the ``weak-valued root-mean-square difference'' between measurement outcomes of the WP$_A$(WP$_B$) and the MA (postmeasurement of B).
%%%%%%%%%%
\begin{equation}
    \label{weak_value}
    \begin{aligned}
       \epsilon(A)^2 &= \sum_{i,f}(a_i-a_f)^2P_{wv}(a_i,a_f), \\
        \eta(B)^2 &= \sum_{i,f}(b_i-b_f)^2P_{wv}(b_i,b_f),
    \end{aligned}
\end{equation}
%%%%%%%%%%
where $P_{wv}(a_i,a_f)$ is the weak-valued joint probability distribution taking the outcomes $a_i$ in the WP$_A$ and $a_f$ in the MA. As described later, we can experimentally estimate $P_{wv}(a_i,a_f)$, and thus $ \epsilon(A)$, by evaluating the probability distribution $P(a_i, a_f)$ that we take the outcomes $a_i$ and $a_f$. Similarly, $\eta(B)$ is given by $P_{wv}(b_i,b_f)$ by taking outcomes $b_i$ in the WP$_B$ and $b_f$ in the postmeasurement of $B$.

In this paper, we report the experimental test of EDRs for a superconducting qubit, placed in a NISQ processor using the weak-probe method. 
One of the advantages of using superconducting qubits over photonic qubits \cite{PhysRevLett.112.020402} is that it's easier to create interactions between multiple qubits. 
Therefore we can assign independent qubits for system, probe $A$, probe $B$, meter and all measurement outcomes to derive error and disturbance can be obtained simultaneously from a single configuration of the quantum circuit. 
Note that WP$_A$ and WP$_B$ were applied only one at a time in Ref. \cite{PhysRevLett.112.020402} since same probe qubit (encoded on single photon's path) is shared for both WP$_A$ and WP$_B$ to measure the system (polarization) state of the same photon. 
%%%%%%%%%%%%%%%%%%%%%%%%%%
\section{Experiment\label{sec:experiment}}
%%%%%%%%%%%%%%%%%%%%%%%%%%
%%%%%%%%%%%%%%%%%%%%%%%%%%
\begin{figure*}[htbp]%fig2
\centering
\includegraphics[width=0.99\textwidth]{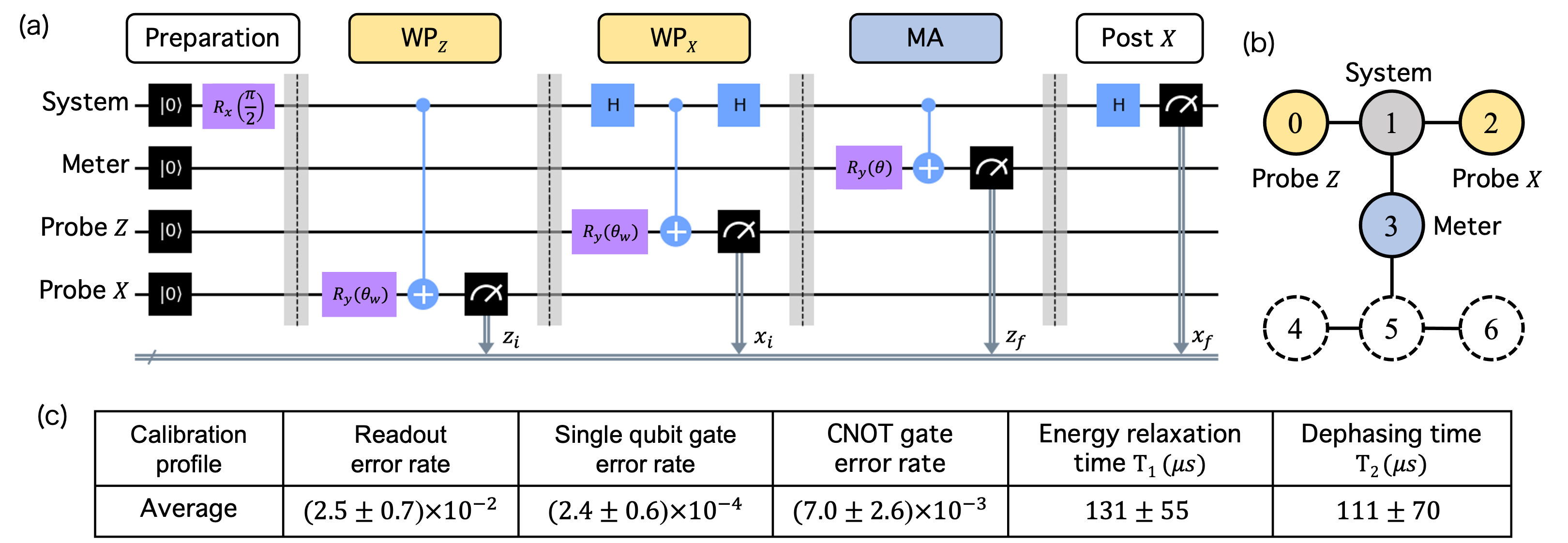} 
\caption{(a) Quantum circuit model of testing EDRs for single-qubit observables $A=Z$ and $B=X$. Our implementation is separated into the state preparation, weak probe $Z$ (WP$_Z$),  weak probe $X$ (WP$_X$), measurement apparatus (MA), and post $X$ measurement. The system qubit prepared in $|R\rangle =( |0\rangle-i|1\rangle)\sqrt{2}$, the eigenstate of $Y$, is indirectly measured by WP$_Z$, WP$_X$, MA, and Post $X$ measurement is followed. The error (disturbance) is given by the “weak-valued root-mean-square difference” between measurement outcomes of the WP$_Z$ and the MA (WP$_X$ and post $X$ measurement). In the experiment, the measurement strength of both WP is set to $\cos\theta_{w}=0.05$ and the measurement strength of the MA, $\cos \theta$, was varied from 0 (no measurement) to 1 (projection measurement). (b) Illustration of qubit assignment on the 7-qubit IBM-Q Falcon processor for the experimental test of EDRs. The system qubit interact independently with their neighbors (Probe $Z$, Probe $X$, Meter) in turn. (c) Calibration profile (averaged over 4 qubits) provided by the IBM-Q processor.} 
\label{q.circuit} 
\end{figure*}
%%%%%%%%%%%%%%%%%%%%%%%%%%
Our implementation is based on the quantum circuit depicted in Fig. \ref{q.circuit}(a). We take the system observable to be measured as $A=Z$ and $B=X$, where $X$, $Y$, and $Z$ denote the Pauli matrices, and $\{|0\rangle, |1\rangle\}$ are eigenbasis of $Z$ with the eigenvalues of \{1, –1\}.
We define weak probe observables in WP$_Z$ and WP$_X$ as $Z_i$ and $X_i$, respectively.  
Meter observable in MA and post $X$ measurement observable for the system are defined as $Z_f$ and $X_f$, respectively. 
We then use the following notation as the measurement outcomes: $a_{i, f}=z_{i, f} =\pm 1$ and $b_{i, f}=x_{i, f} =\pm 1$.

We employ three cascaded circuits as WP$_Z$, WP$_X$, and MA. All three circuits work in the same manner. 
In the MA, meter qubit initialized to $|0\rangle$ is rotated by 
%%%%%%%%%%
$R_{y}(\theta)=\exp(-i \frac{\theta}{2} Y ) = 
\begin{pmatrix} 
\cos\frac{\theta}{2} & -\sin\frac{\theta}{2}\\ 
\sin\frac{\theta}{2}  &  \cos\frac{\theta}{2}
\end{pmatrix}$,
%%%%%%%%%%
%where $0\leq\theta\leq \pi/2$.
where $0\leq\theta\leq \frac{\pi}{2}$.
Then, the meter qubit is subjected to a controlled-NOT (CNOT) operation with the system qubit. 
The positive operator valued measure (POVM) elements corresponding to the outcomes of $z_f=\pm1$ are
%%%%%%%%%%
\begin{equation}
    \label{eq:POVM}
    \begin{aligned}
        \Pi_{z_f=\pm 1} = \frac{1}{2}(I \pm (\cos\theta) Z).
    \end{aligned}
\end{equation}
%%%%%%%%%%
Here $\cos\theta$ is the measurement strength of the MA. 
By changing $\cos\theta$  from 0 to 1, $ \Pi_{z_f=\pm 1} $ change from identity (no measurement) to projector (strong measurement). The WP works in exactly the same manner as the MA except that the measurement strength of the WP is $\cos\theta_w$. In order to keep the WP’s measurement strength sufficiently weak, $\theta_w$ should be close to $\pi/2$. In addition, two Hadamard gates ($H$) are inserted to the system qubit before and after the CNOT in the WP$_X$, where weak measurement for $X$ is taken.

To take the most stringent test of the EDRs, we chose the system state as $|R\rangle =( |0\rangle-i|1\rangle)\sqrt{2}$, an eigenstate of $Y$, so that the rhs of the EDRs become the maximum value in the qubit measurement: $C=|\langle[Z, X]\rangle|/2 = |\langle Y\rangle| = 1$. 
To prepare the system state in $|R\rangle$, as shown in the preparation part of Fig. \ref{q.circuit}(a), the system qubit initialized to $|0\rangle$ is rotated by a $R_{x}(\theta=\pi/2)$ gate, 
where 
%%%%%%%%%%
$R_{x}(\theta)=\exp(-i \frac{\theta}{2} X ) = 
\begin{pmatrix} 
\cos\frac{\theta}{2} & -i\sin\frac{\theta}{2}\\ 
-i\sin\frac{\theta}{2} &  \cos\frac{\theta}{2}
\end{pmatrix}$.
%%%%%%%%%%
A meter for MA and weak probes for WP$_Z$ and WP$_X$ are initialized to $|0\rangle$, interact with system qubit in sequence, and finally measured in $\{|0\rangle, |1\rangle\}$ basis.

In the experiment, the measurement strength for both WP$_Z$ and WP$_X$ was set to $\cos \theta_{w} = 0.05$ that produced a very small disturbance in the initial state. We expect the rhs of EDRs, $C = 4/ (3+\cos2\theta_w)-1=0.995$, which is closed to the ideal value 1.
%$C = \frac{4}{3+\cos2\theta_w}-1=0.995$
Then, the signal photon was subjected to the MA and post $X$ measurement, consisting of a Hadamard gate and projection on $Z$ basis, where measurement basis is switched from $Z$ to $X$.
From Eq. (\ref{weak_value}) and the expression of weak-valued joint probability distribution \cite{Lund_2010}, the error and disturbance are given by
%%%%%%%%%%
\begin{equation}
    \label{err_disturbance2}
    \begin{aligned}
        \epsilon(Z)^2 &= 2\left( 1- \frac{1}{\cos\theta_w/2}\sum_{z_i,z_f}z_i z_f P(z_i, z_f)\right),\\
        \eta(X)^2 &= 2\left(1- \frac{1}{\cos\theta_w/2}\sum_{x_i,x_f}x_i x_f P(x_i, x_f)\right), 
    \end{aligned}
\end{equation}
%%%%%%%%%%
where the subscript $z_i$, $z_f$, $x_i$, $x_f$ denotes the outcomes of the WP$_Z$, MA, WP$_X$, and post $X$ measurement, respectively. 
By employing this quantum circuit, both the error and disturbance can be experimentally quantified simultaneously.

Experiments were conducted on the 7-qubit IBM Quantum Falcon Processor (IBM\_perth) \cite{ibm2021} under the open-source framework Qiskit. Fig. \ref{q.circuit}(b) illustrates the qubit assignments for the EDR test.
Each node (circle) represents a superconducting qubit and CNOT operations can be performed between two qubits connected by a line. 
In our implementation, qubit 1 surrounded by three neighboring qubits is assigned as the system qubit and qubit 0, 2, 3 are assigned as probe $Z$, probe $X$, and meter, respectively.
As our EDR test presented in Fig. \ref{q.circuit}(a) only requires 4 qubits, qubit 4, 5, and 6 were not used. 
Note that we can apply CNOT operations between any two qubits of 7 qubits (even between non-neighboring qubits) after consecutive swap operations which would introduce additional gate errors.
%%%%%%%%%%%%%%%%%%%%%%%%%%
\section{Results\label{sec:results}}
%%%%%%%%%%%%%%%%%%%%%%%%%%
%%%%%%%%%%%%%%%%%%%%%%%%%%
\begin{figure}[t] %fig3
\centering
\includegraphics[width=0.48\textwidth]{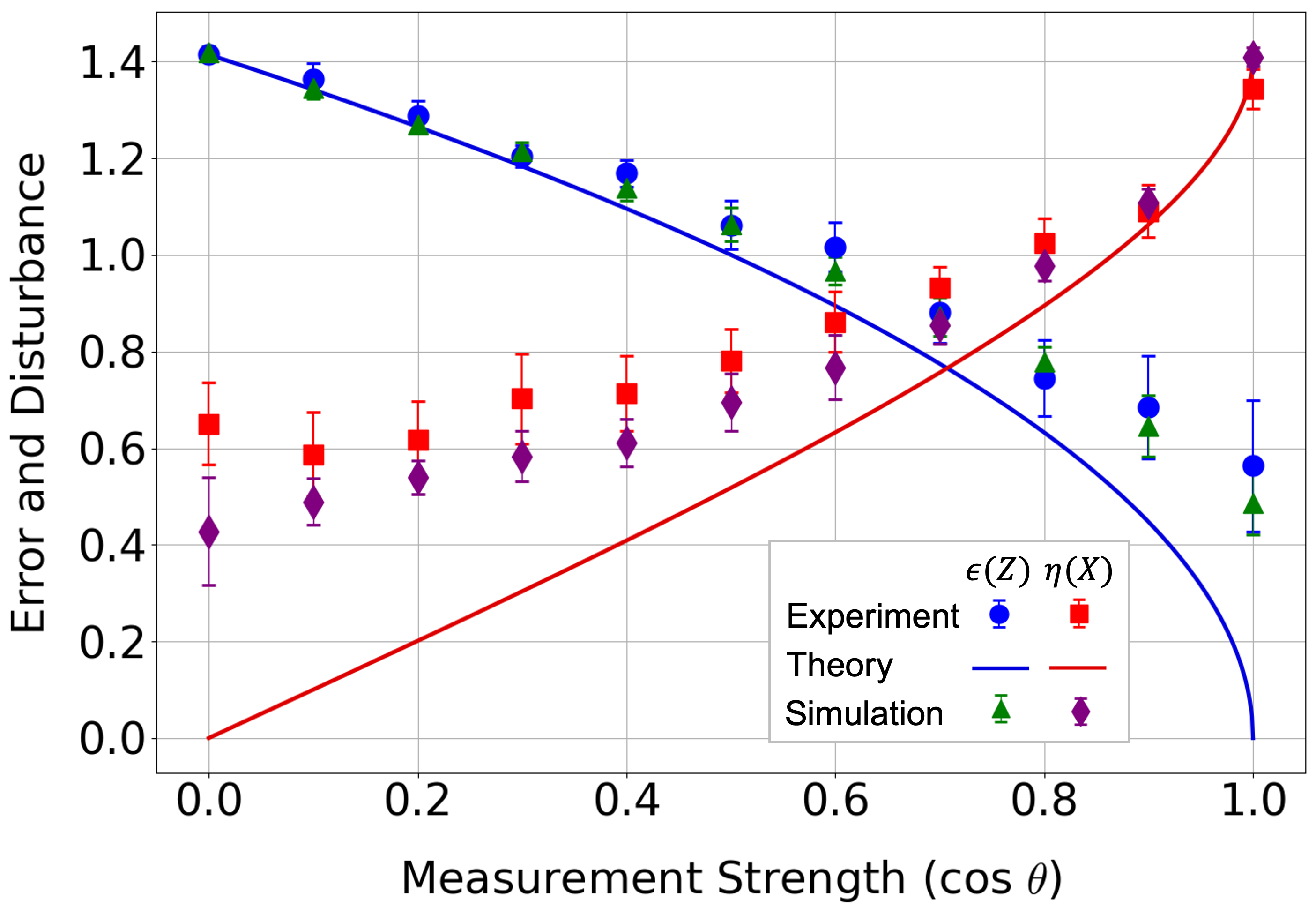} 
\caption{Experimentally obtained error $\epsilon(Z)$ (blue circles) and disturbance $\eta(X)$ (red squares) as functions of the measurement strength $\cos\theta$. Solid curves are the theoretically calculated error and disturbance for perfect implementation of the quantum circuit presented in Fig. \ref{q.circuit}. Classically simulated error (green triangles) and disturbance (purple diamonds) are obtained from IBM open-source framework Qiskit after noise in the quantum processor is taken into account. See text for details.} 
\label{err_dis} 
\end{figure}
%%%%%%%%%%%%%%%%%%%%%%%%%%
To evaluate $\epsilon(Z)$ and $\eta(X)$ using Eq. (\ref{err_disturbance2}), we experimentally obtain $P(z_i, z_f)$ and $P(x_i, x_f)$ by executing the designed quantum circuit in Fig. \ref{q.circuit}(a) on the  IBM-Q processor 100,000 times and analyzing the statistics of the measurement outcomes $\{z_i, x_i, z_f, x_f\}$. The quantities $\epsilon(Z)$ and $\eta(X)$ averaged for 10 repeated measurements, $10^5\times10=10^6$ measurements in total, are shown in Fig. \ref{err_dis} throughout the range of the measurement strength from no measurement ($\cos\theta=0$) to projective measurement ($\cos\theta=1$). 
The error bars represent the rms error of 10 repeated measurements. 
The experiment runs approximately in 40 min in total.
Solid curves are the theoretically calculated error (blue) and disturbance (red) for perfect implementation of the quantum circuit presented in Fig. \ref{q.circuit}(a). 
Fig. \ref{err_dis} also shows the simulated error (green triangles) and disturbance (purple diamonds), obtained from IBM open-source framework Qiskit after the calibration data of the quantum processor is taken into account. 
The calibration data contains full information characterizing the quantum processor such as gate error rates and coherence times. 
Table in Fig. \ref{q.circuit}(c) lists the calibration data (averaged over four qubits) that are expected to have a major impact on the simulation. 

We clearly see the trade-off relation between the error and disturbance: as the measurement strength increases, $\epsilon(Z)$ decreases while $\eta(X)$ increases. 
The presence of unavoidable noise in the quantum processor causes deviation of both the simulation results and the experimental data from the ideal theoretical expectations. 
Particularly, when the measurement strength is gradually reduced to $\cos \theta = 0$, ideally we expect no disturbance, i.e., $\eta(X)=0$, but the measured disturbance is saturated around $\eta(X) = 0.6$ due to intrinsic noisy property of the NISQ device. In the same way, for the region measurement strength reach to $\cos\theta = 1$, the measurement error $\epsilon(Z)$ is ended up around $0.4$ instead of 0. 
The experimentally measured error and disturbance not perfectly but closely follow the theoretically simulated error and disturbance in which the calibration data is taken into account.
The deviation from the simulation might originate from additional experimental imperfections and limitations of simulations that cannot account for all real-time noise of the quantum processor. 
%%%%%%%%%%%%%%%%%%%%%%%%%%
\begin{figure}[t]%fig4
\centering
\includegraphics[width=0.48\textwidth]{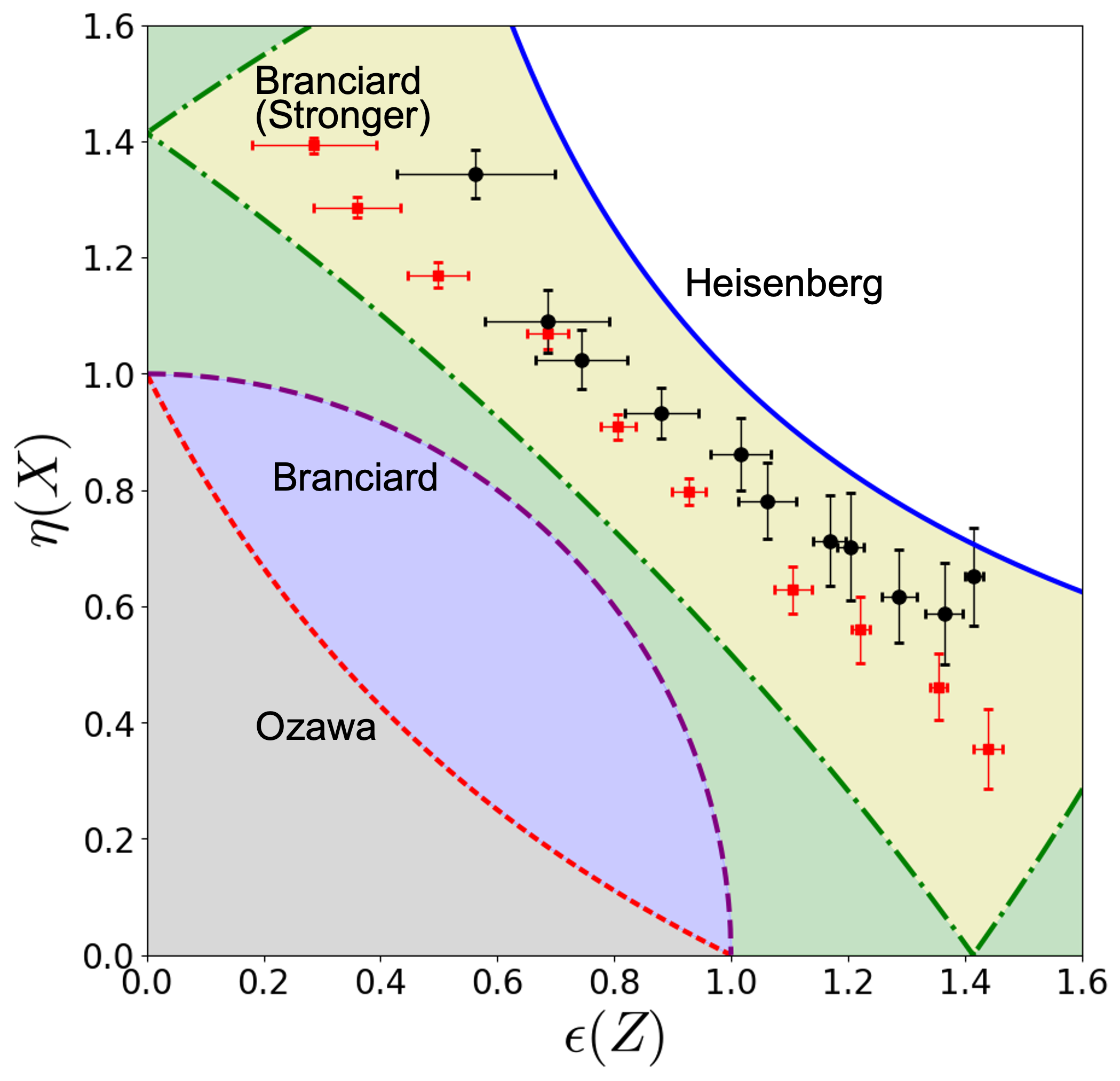} 
\caption{Comparison of EDRs’ lower bounds in the error–disturbance plot. Blue (solid) curve: the Heisenberg bound in Eq. (\ref{H_edr}). Red (short dashed) curve: the Ozawa bound in Eq. (\ref{O_edr}). Purple (long dashed) curve: the Branciard bound in Eq. (\ref{B_edr}). Green (dotted-dashed) curve: the stronger Branciard bound in Eq. (\ref{B_edr2}). Black circles: experimental data shown in Fig. \ref{err_dis}. Red squares: experimental data obtained in EDR test for polarization measurement of single photons \cite{PhysRevLett.112.020402}. The lower-left side of each bound is the forbidden region by the corresponding EDR. Each bound was calculated for C=1.} 
\label{edr_lhs} 
 \end{figure}

In Fig. \ref{edr_lhs}, we plot the predicted lower bounds of the EDRs in Eqs. (\ref{H_edr}), (\ref{O_edr}), and (\ref{B_edr}), together with the experimental data. 
We also plot the stronger Branciard EDR (green) that is applicable to the case (including ours) where the system and meter observables are both $\pm1$ valued and $\langle A \rangle = \langle B \rangle = 0$ (hence $\sigma(A)=\sigma(B)=1$) \cite{branciard2013error}
%%%%%%%%%%
\begin{equation}
\label{B_edr2}
\left[ \tilde{\epsilon}(A)^2+\tilde{\eta}(B)^2+2\tilde{\epsilon}(A)\tilde{\eta}(B)\sqrt{1-C^2}\right]^{1/2}  \geq C,
\end{equation}
%%%%%%%%%%
where $\tilde{\epsilon}=\epsilon\sqrt{1-\epsilon^2/4}$ and $\tilde{\eta}=\eta\sqrt{1-\eta^2/4}$.
Under the Heisenberg EDR, the error or disturbance must be infinite when the other goes to zero, while other EDRs allow a finite error or disturbance even when the other is zero. 
The lower-left side of each bound is the forbidden region by the corresponding EDR. Each bound was calculated for $C=1$. 

We clearly see that the experimental data (black circles) is present in the lower-left side of Heisenberg EDR and right-upper side of other EDRs.  
Therefore, our experimental results demonstrate the clear violation of the Heisenberg EDR, while the Ozawa and Branciard EDRs are always satisfied throughout the full range of measurement strength. 
We see that the Branciard EDRs are stronger than the Ozawa EDR. 
Our experimental data were closer to the stronger Branciard bound (dotted-dashed curve) given in Eq. (\ref{B_edr2}), which could be saturated by ideal experiments \cite{PhysRevLett.112.020402}, than the other EDR bounds. 
Note that we need to keep sufficiently weak measurement strengths for WP$_Z$ and WP$_X$ ($\cos\theta_w=0.05$ in our case) to observe validity of Ozawa and Branciard EDRs.

%%%%%%%%%%%%%
% photonic qubit vs SC qubits in NISQ
%%%%%%%%%%%%%
For the comparison, we also plot experimental data (red squares) obtained in the investigations of EDRs for polarization measurement of single photons \cite{PhysRevLett.112.020402}.
We can see that our data measured for superconducting qubits is further deviated from the stronger Branciard bound than the measurement data for single photons.
Our data also shows slightly larger error bars compared to the photonic implementation with a comparable sampling number due to intrinsic noisy property of NISQ device. 
By comparing the two data sets, we can see that the NISQ device introduces additional noise and errors along with the benefit of introducing multi-qubit interactions easily. 
This noise will introduce imperfect control over qubits and will place limitations on realizing the designed quantum measurement protocols in the quantum processor. 
Therefore, developing and leveraging tools for characterizing noise and error, mitigating them, and verifying quantum processing will be crucial in implementing the measurement protocols on quantum devices and for further applications \cite{PhysRevA.100.052315, PhysRevA.103.042603}.

\section{Conclusion\label{sec:conclusion}}
We have experimentally tested the Heisenberg, Ozawa, and Branciard EDRs  in generalized, variable strength measurements of superconducting qubits on a NISQ processor for the first time. 
Making use of sequential weak measurements that keeps the initial signal state practically unchanged prior to the main measurement, we demonstrate that the Heisenberg EDR is violated, yet the Ozawa and Branciard EDRs are valid throughout the range of measurement strengths from no measurement to projection measurement. 
In particular, our results were closer to the stronger Branciard bound (dotted-dashed curve) given in Eq. (8), than the other EDR bounds, similar to what was seen in the photonic implementation in \cite{PhysRevLett.112.020402}.
Such experimental investigation of the EDRs will be of demanded importance not only in understanding fundamentals of physical measurement but also in developing novel measurement protocols for quantum information and communications. 

It is important to acknowledge that the presence of intrinsic noise of NISQ system causes deviations between the experimental results and the theoretical expectation, which cannot be fully compensated for even with extensive sampling to reduce the statistical error. 
Nevertheless, the NISQ processor proves its prospect as a testbed of new quantum tasks, by performing proof-of-principle demonstration of quantum measurement protocols and providing results that tend to follow the theoretical predictions.
Particularly, the capability of introducing multi-qubit entangling operations of NISQ system would allow experimental investigations of various multi-qubit measurement protocols \cite{edamatsu2016complete, Vidil_2021, 10.1038/s41534-023-00688-7}.
By addressing and exploring the existing challenges posed by noise and errors in NISQ systems \cite{PhysRevA.100.052315, PhysRevA.103.042603}, we can further enhance the fidelity and reliability of quantum experiments and pave the way for future advancements in quantum technologies.

\begin{acknowledgments}
This research was supported by MEXT Quantum Leap Flagship Program (MEXT Q-LEAP) Grant No. JPMXS0118067581. 
T.D. was supported by JST, the establishment of university fellowships towards the creation of science technology innovation Grant No. JPMJFS2102.
We acknowledge the use of IBM Quantum services for this work. 
The views expressed are those of the authors, and do not reflect the official policy or position of IBM or the IBM Quantum team.

T.D. and S.B. contributed equally to this work.
\end{acknowledgments}

\bibliography{EDR_TD_SB}
\end{document}